\documentclass[twocolumn,aps,pra,showpacs,floatfix,groupeaddress]{revtex4}

\usepackage{amsmath}
\usepackage{amsfonts}
\usepackage{amssymb}
\usepackage{bm}
\usepackage{bbm}
\usepackage{graphicx}
\usepackage{color}
\usepackage{maybemath}
\usepackage{fancybox}

\newcommand{\dd}{\mathrm{d}}
\newcommand{\ee}{\mathrm{e}}
\newcommand{\ii}{\mathrm{i}}

\newcommand{\plus}{{\mbox{{\bf{\tiny +}}}}}

\newcommand{\wtilde}[1]{\mbox{$\widetilde #1$}}

\begin{document}

%
%
\title{Gravitationally Coupled Dirac Equation for Antimatter}

\author{U. D. Jentschura}
\affiliation{Department of Physics,
Missouri University of Science and Technology,
Rolla, Missouri 65409, USA}
\affiliation{MTA--DE Particle Physics Research Group,
P.O.Box 51, H--4001 Debrecen, Hungary}

\begin{abstract}
The coupling of antimatter to gravity is of general interest because of
conceivable cosmological consequences (``surprises'') related to dark energy
and the cosmological constant. Here, we revisit the derivation of the 
gravitationally coupled Dirac equation and 
find that the prefactor of a result given previously in
[D.~R. Brill and J.~A. Wheeler, Rev. Mod. Phys. {\bf 29}, 465 (1957)]
for the affine connection matrix is in need of a correction.
We also discuss the conversion of the curved-space 
Dirac equation from ``East--Coast'' to ``West--Coast'' conventions, 
in order to bring the gravitationally coupled
Dirac equation to a form where it can easily be unified 
with the electromagnetic coupling as it is commonly used in modern particle 
physics calculations. The Dirac equation describes anti-particles as
negative-energy states.
We find a symmetry of the gravitationally 
coupled Dirac equation, which connects particle and 
antiparticle solutions for a general space-time metric 
of the Schwarzschild type and implies that particles and 
antiparticles experience the same coupling to the 
gravitational field, including all relativistic quantum
corrections of motion. Our results demonstrate the consistency of quantum
mechanics with general relativity and imply that a conceivable difference of
gravitational interaction of hydrogen and antihydrogen should directly be
attributed to a a ``fifth force'' (``quintessence'').
\end{abstract}


\maketitle

%
%
\section{Introduction}
\label{sec1}

In view of the recent dramatic progress of antimatter gravity
experiments~\cite{GaEtAl2008,AmEtAl2011}, it seems indicated to 
reexamine the theoretical status of antimatter coupling to gravity.
A number of experimental
collaborations are actively pursuing related
experiments~\cite{ALPHA,ATHENA,ASACUSA,ATRAP,AGELOI,Ke2008}. 
A key factor in recent experimental 
progress~\cite{AmEtAl2011} of the ALPHA collaboration
has been their special Penning--Ioff\'{e} trap which
simultaneously traps both positrons as well as 
antiprotons. Superimposed on the Penning trap fields
(which trap the charged constituent particles),
the antihydrogen atom Ioff\'{e}
trap employed by ALPHA relies on a strong octupole magnetic field 
configuration generated by eight superconducting  current bars,
which wind back on themselves in a sinuous pattern,
glued to the inner chamber of the ALPHA 
experiment by a three-dimensional winding machine
at Brookhaven National Laboratory (see Ref.~\cite{ALPHACOILS}).
This leads to an effective trapping of positrons and antiprotons,
and antihydrogen atoms.

Antimatter gravity experiments aim to test the 
interaction of antihydrogen atoms with gravitational fields.
According to general relativity~\cite{MiThWh1973},
gravitational interactions can be described by the 
induced space-time curvature around massive objects. 
Furthermore, on the classical level, the motion of a 
particle in curved space-time is described 
by the following geodesic equation~\cite{MiThWh1973}
\begin{equation}
\label{eqEP}
\frac{\dd^2 x^\mu}{\dd^2 s} +
{\Gamma^\mu}_{\rho\sigma} \, \frac{\dd x^\rho}{\dd s} \,
\frac{\dd x^\sigma}{\dd s} = 0 \,,
\end{equation}
which implies that a particle of mass $m$ experiences a ``force''
$F^\mu = m \, \dd^2 x^\mu/\dd s^2$ and moves along a 
zero geodesic in the gravitationally curved space-time
($s$ is the proper time).
Here, the ${\Gamma^\mu}_{\rho\sigma}$ are the Christoffel 
symbols~\cite{MiThWh1973},
derived from the curved-space metric $\overline g_{\mu\nu}$ as follows,
\begin{equation}
\label{chr}
\Gamma_{\alpha\rho\sigma} = \frac12 \, \left(
\frac{\partial \overline g_{\alpha\sigma}}{\partial x^\rho} +
\frac{\partial \overline g_{\alpha\rho}}{\partial x^\sigma} -
\frac{\partial \overline g_{\rho\sigma}}{\partial x^\alpha} 
\right) \,,
\end{equation}
where the Einstein summation convention is used.
The ${\Gamma^\mu}_{\rho\sigma}$ are derived from the 
$\Gamma_{\alpha\rho\sigma}$ by raising the first index with the 
help of the metric, i.e., ${\Gamma^\mu}_{\rho\sigma} = 
\overline g^{\mu\alpha} \, \Gamma_{\alpha\rho\sigma}$. We should clarify that in the 
current article, in a somewhat non-standard
notation, the symbol $\wtilde{g}_{\mu\nu}$ will be reserved for the 
flat-space metric in the following, whereas
$\overline g^{\mu\alpha}$ denotes the metric of curved space.
If, according to Einstein's equivalence principle,
we assume that gravitational mass and inertial 
mass are proportional to each other, then classical 
geometrodynamics~\cite{MiThWh1973},
on the basis of Eq.~\eqref{chr},
makes the unique prediction that the force on a
particle and antiparticle in a gravitational field are the same,
provided the mass of particle and antiparticle are equal,
i.e., both particle as well as antiparticle motion 
are described by Eq.~\eqref{eqEP}.
However, on the quantum level, the 
situation is less clear.
It is often argued~\cite{AGELOI} that ``general relativity 
is incompatible with quantum mechanics'' and 
that, assuming rather peculiar couplings of antimatter
to gravity~\cite{Ko1996}, one can imagine that 
antimatter actually is repulsed by gravity.
This observation provides part of the motivation for a number of 
antimatter gravity experiments currently under 
preparation~\cite{ALPHA,ATHENA,ASACUSA,ATRAP,AGELOI,Ke2008}.

Here, we reexamine the status of theoretical predictions regarding
the coupling of Dirac particles and antiparticles to curved space-time.
Indeed, closer inspection shows that considerable insight into the
gravitational coupling of antiparticles can be gained based on rather
straightforward generalizations of previous treatments which rely on a
combination of relativistic quantum mechanics with general relativity.  We note
the works of Brill and Wheeler~\cite{BrWh1957}, Boulware~\cite{Bo1975prd}, and
Soffel, M\"{u}ller and Greiner~\cite{SoMuGr1977}.  The Dirac
equation~\cite{Di1928a,Di1928b} describes both particles and antiparticles
simultaneously, and symmetries of the solutions which connect particles and 
antiparticles are therefore relevant for antigravity experiments.
We find that it is highly indicated to 
revisit a number of aspects of the derivation.
We employ units with $\hbar = c = \epsilon_0 = 1$.

%
%
\section{Formalism}
\label{sec2}

Antihydrogen consists of two spin-$1/2$ particles, the electron and the proton.
Spin-$1/2$ particles
are described by the Dirac equation.  In curved and flat space-time, 
respectively, the anticommutators
$\{ \cdot, \cdot \}$ of the Dirac $\gamma$ matrices fulfill the algebraic
relations 
\begin{equation}
\label{gmunu}
\{ \overline \gamma^\mu(x), \overline \gamma^\nu(x) \} = 
2 \, \overline g^{\mu\nu}(x) \,,
\qquad
\{ \wtilde{\gamma}^\mu, \wtilde{\gamma}^\nu \} = 
2 \, \wtilde g^{\mu\nu} \,,
\end{equation}
where the curved-space metric is 
$\overline g^{\mu\nu}$ with $\mu,\nu =0,1,2,3$, while the 
local flat-space metric (``vierbein'') in our conventions 
is $\wtilde g^{\mu\nu} = {\rm diag}(1,-1,-1,-1)$.
The precise form of the $\overline \gamma^\mu(x)$ matrices
depends on the space-time geometry and in particular, 
on the space-time coordinate $x$.
We use the ``West-Coast'' signature ${\rm diag}(1,-1,-1,-1)$ for the 
free-space metric $\wtilde g$ instead of the 
``East--Coast'' conventions ${\rm diag}(1,1,1,-1)$
or ${\rm diag}(-1,1,1,1)$, in order to ensure compatibility with the 
sign convention usually adopted in the modern
particle physics literature~\cite{BjDr1964,BjDr1965,ItZu1980,PeSc1995}.

This is different from the sign conventions adopted 
in the traditional literature on general relativity~\cite{MiThWh1973},
and also different from the sign conventions
used in previous works on the gravitationally 
coupled Dirac equation~\cite{BrWh1957,Bo1975prd,SoMuGr1977}.
E.g., the conventions of Misner, Thorne and Wheeler~\cite{MiThWh1973}
are given in Eq.~(2.10) on page~53 of Ref.~\cite{MiThWh1973}
and involve a metric ${\rm diag}(-1,1,1,1)$.
with $\mu,\nu =0,1,2,3$. If we ever wish to study the combined 
``gravito-magnetic'' effect of gravitational
and electromagnetic fields on Dirac particles simultaneously,
and conceivably use established results for the
electromagnetic sector in a perturbative, then it is helpful
to convert the gravitational Dirac equation 
into ``West--Coast'' conventions, because these
are used in the particle physics and quantum electrodynamics
(QED) literature.

One might ask why we are using the tilde in order to denote the 
flat-space metric, not just the plain symbol $g_{\mu\nu}$.
The answer to that question is as follows.
We would like to be as unique in our notation as possible,
and avoid possible confusion upon comparison with 
the literature~\cite{BrWh1957,Bo1975prd,SoMuGr1977}.
In Refs.~\cite{BrWh1957,Bo1975prd,SoMuGr1977}, the curved-space
Dirac matrices are denoted as $\gamma^\mu$, but in the 
particle physics literature~\cite{BjDr1964,BjDr1965,ItZu1980,PeSc1995},
one denotes the flat-space matrices as $\gamma^\mu$.
There is no way to unify the notations without introducing 
some ambiguity, and we have therefore decided to 
differentiate the matrices either by overlining or using the 
tilde, making their identification unique.

The flat-space action for the free Dirac particle in special relativity 
reads as 
\begin{align}
\label{S0}
S_0 =& \; \int \dd^4 x \, \overline \psi(x) \,
\left( \ii \, \wtilde\gamma^\mu \partial_\mu - m \right) \psi(x) 
\nonumber\\[0.007ex]
=& \; \int \dd^4 x \, \overline \psi(x) \,
\left( \frac{\ii}{2} \, 
\wtilde\gamma^\mu \overleftrightarrow{\partial}_\mu - m \right) \psi(x) \,,
\end{align}
where $\overline \psi(x) = \psi(x)^\plus \, \wtilde{a}$ is the 
Dirac adjoint, 
$\partial_\mu = \partial/\partial x^\mu$ is 
the derivative with respect to $x^\mu$,
and the symmetric derivative operator acts as 
\begin{equation}
A(x) \, \overleftrightarrow{\partial}_\mu \, B(x) \equiv
A(x) \, \partial_\mu \, B(x) - B(x) \, \partial_\mu \, A(x) \,.
\end{equation}
Furthermore, $\wtilde{a}$ is a Hermitizing matrix 
with the property 
\begin{equation}
\wtilde{a} \, \left( \wtilde{\gamma}^\mu \right)^\plus \, \wtilde{a} = 
\wtilde{\gamma}^\mu \,.
\end{equation}
Here, $b^\plus$ denotes the Hermitian conjugate of a matrix $b$.
An infinitesimal global Lorentz transformation $\Lambda$ and the corresponding 
spinor Lorentz transformation $S(\Lambda)$ 
in flat space then read as
\begin{subequations}
\label{LL}
\begin{align}
{\Lambda^\mu}_\nu =& \; {{\mbox{$\wtilde g$}}^\mu}_\nu + 
{\wtilde \omega^\mu}_\nu \,,
\\[0.007ex]
S(\Lambda) =& \; 1 -\frac{\ii}{4} \, 
\wtilde{\sigma}^{\alpha\beta} \, 
\wtilde{\omega}_{\alpha\beta} \,,
\\[0.007ex]
\wtilde{\sigma}^{\alpha\beta} =& \;
\frac{\ii}{2} \, \left[ \wtilde\gamma^\alpha, 
\wtilde\gamma^\beta \right] \,.
\end{align}
\end{subequations}
Here, ${\wtilde{\omega}^\mu}_\nu + {\wtilde \omega_\nu}^\mu = 0$.
In formulating the generators $\wtilde{\sigma}^{\alpha\beta}$
of spinor Lorentz transformations, we follow the conventions of
Chap.~2 of Ref.~\cite{ItZu1980}. Furthermore, in view of the relation
\begin{equation}
\label{C}
\left[ \wtilde \gamma^\mu, 
\wtilde \sigma^{\alpha \beta} \right] =
2 \, \ii \, \wtilde g^{\mu\alpha} \, \wtilde \gamma^\beta - 
2 \, \ii \, \wtilde g^{\mu\beta} \, \wtilde \gamma^\alpha \,,
\end{equation}
the $\wtilde \gamma$ matrices are shape-invariant 
under Lorentz transformations,
\begin{equation}
{\wtilde \gamma}'^\mu = {\Lambda^\mu}_\nu  \, S(\Lambda) \, 
{\wtilde \gamma}^\nu \, S(\Lambda)^{-1} = {\wtilde \gamma}^\mu \,,
\end{equation}
and $\overline \psi$ transforms with the inverse Lorentz
transformation,
$\overline\psi'(x') = \overline\psi(x) \,  S(\Lambda)^{-1}$.
This can be shown easily by observing that
\begin{equation}
\wtilde a \, \left( S(\Lambda) \right)^\plus \, 
\wtilde a = S(\Lambda)^{-1} \,.
\end{equation}
Standard representations of the flat-space 
Dirac matrices $\wtilde \gamma$ include the Dirac
and the Majorana representation~\cite{ItZu1980,PeSc1995}.

The generalization of the Dirac action~\eqref{S0} 
to curved space-time involves two steps: (i)~an obvious 
generalization of the anticommutator relations~\eqref{gmunu}
to curved space,
$\{ \overline\gamma^\mu(x), \overline\gamma^\mu(x) \} = 
2 \, \overline g^{\mu\nu}(x)$,
and (ii)~a coupling of the derivative operator $\partial_\mu$
in the Dirac equation to the gravitational field on the basis of 
a covariant derivative,
in the sense of the replacement 
$\partial_\mu \to \nabla_\mu = \partial_\mu - \Gamma_\mu$, where 
$\nabla_\mu$ is the covariant derivative and 
$\Gamma_\mu \equiv \Gamma_\mu(x)$ is the affine connection matrix.
The action for the Dirac particle in curved space-time 
then reads as
\begin{equation}
\label{Sdirac}
S = \int \dd^4 x \, \sqrt{-\det \, \overline g} \;\; \overline\psi(x) \;
\left( \ii \, \overline\gamma^\mu(x) \nabla_\mu - m \right) \;
\psi(x) \,,
\end{equation}
where $\det \, \overline g = {\rm det} \, \overline g_{\mu\nu} < 0$ is the 
determinant of the space-time metric,
and $\overline \psi = \psi^+ \, {\overline a}(x)$ is the 
curved-space Dirac adjoint, where ${\overline a}(x)$ is a Hermitizing 
matrix with the local properties
\begin{align}
\label{alocal}
{\overline a}(x) \, \left( \overline\gamma^\mu(x) \right)^\plus \, 
{\overline a}(x) =& \; \overline\gamma^\mu(x) \,,
\\[0.007ex]
{\overline a}(x) \, \left( S(L(x)) \right)^\plus \, 
{\overline a}(x) =& \; S(L(x))^{-1} \,.
\end{align}
Here, $S(L(x))$ is the spinor transformation 
corresponding to an infinitesimal,
local Lorentz transformation $L(x)$ and reads as
\begin{subequations}
\label{Lloc}
\begin{align}
{L(x)^\mu}_\nu =& \; {{\mbox{$\overline g$}}^\mu}_\nu + 
{\overline \omega^\mu}_\nu(x) \,,
\\[0.007ex]
S(L(x)) =& \; 1 -\frac{\ii}{4} \;
\overline{\sigma}^{\alpha\beta}(x) \;
\overline{\omega}_{\alpha\beta}(x) \,,
\\[0.007ex]
\overline{\sigma}^{\alpha\beta}(x) =& \;
\frac{\ii}{2} \, \left[ \overline\gamma^\alpha(x), 
\overline\gamma^\beta(x) \right] \,.
\end{align}
\end{subequations}
These equations generalize Eq.~\eqref{LL} to curved space-time
and ensure that the $\overline\gamma$ matrices are shape-invariant
under Lorentz transformations,
\begin{equation}
\label{ensure}
{\overline\gamma}'^\mu(x) = {L(x)^\mu}_\nu  \, S(L(x)) \, 
{\overline\gamma}^\nu(x) \, S(L(x))^{-1} = {\overline\gamma}^\mu(x) \,.
\end{equation}
From now on, we shall suppress the space-time coordinate
argument $x$ in the 
$\overline\gamma$ and $\overline\sigma$ matrices.
The generalization of Eq.~\eqref{C} to general 
relativity is given by 
\begin{equation}
\label{CGR}
\left[ \overline\gamma^\mu, 
\overline \sigma^{\alpha \beta} \right] =
2 \, \ii \, \overline g^{\mu\alpha} \, \overline\gamma^\beta - 
2 \, \ii \, \overline g^{\mu\beta} \, \overline\gamma^\alpha \,.
\end{equation}
One can show this relation using Eq.~\eqref{gmunu} only.
For tensors, the covariant derivative  $\nabla_\mu$ is well established
[see~Exercise~8.4 on page~211 of Ref.~\cite{MiThWh1973}],
but for spinors, a nontrivial generalization is required.
Let us assume the structure
\begin{equation}
\label{ansatzG}
\nabla_\nu \psi = (\partial_\nu - \Gamma_\nu)\psi \,,
\end{equation}
where the affine connection matrix 
$\Gamma_\nu$ remains to be determined.
We postulate that the covariant derivative operator 
commutes with the current matrix $\overline\gamma^\mu(x)$,
i.e., $[\overline\gamma_\mu(x), \, \nabla_\nu] = 0$.
Then,
\begin{equation}
\label{commutes}
\overline\gamma_\mu(x) \, \nabla_\nu \psi(x) = 
\nabla_\nu \, \left( \overline\gamma_\mu(x) \psi(x) \right) \,,
\end{equation}
and we can symmetrize Eq.~\eqref{Sdirac} as follows,
\begin{equation}
\label{symS}
S = \int \dd^4 x \; \sqrt{-\det \overline g} \;\; \overline\psi(x) \, 
\left( \frac{\ii}{2} \overline\gamma^\mu(x) \, 
\overleftrightarrow{\nabla}_\mu - m \right) \, \psi(x) \,.
\end{equation}
By variation, the gravitationally coupled Dirac equation 
is obtained as
\begin{equation}
\label{gravdirac}
\left( \ii \, \overline\gamma^\mu \, \nabla_\mu - m \right) \psi(x) = 0 \,.
\end{equation}
An additional electromagnetic field 
could be incorporated by the replacement
$\nabla_\mu \to \nabla_\mu + i \, q \, A_\mu$, 
where $A_\mu$ is the vector potential
and $q$ is the charge. However, the gravitational Dirac
equation is primarily interesting when all electromagnetic effects
are compensated and the residual gravitational
interaction dominates the kinematics.

The affine connection matrices~$\Gamma_\mu$ remain 
to be determined. Using the ansatz~\eqref{ansatzG},
one can write the condition
given in Eq.~\eqref{commutes} 
as follows~\cite{BrWh1957,Bo1975prd,SoMuGr1977,ShCa1991,ShVa1992}:
\begin{equation}
\label{condition}
\nabla_\nu \overline\gamma_\mu = 
\partial_\nu \overline\gamma_\mu -
{\Gamma^\rho}_{\mu\nu} \overline\gamma_\rho 
- \Gamma_\nu \, \overline\gamma_\mu
+  \overline \gamma_\mu \, \Gamma_\nu = 0 \,.
\end{equation}
The ${\Gamma^\rho}_{\mu\nu} \overline\gamma_\rho $
are the Christoffel symbols defined in Eq.~\eqref{chr}.
For a Lorentz vector $T^\mu$, we recall that~\cite{MiThWh1973}
\begin{subequations}
\begin{align}
\label{TA}
\nabla_\mu T_\alpha =& \; 
\partial_\mu T_\alpha - {\Gamma^\lambda}_{\alpha\mu} \, T_\lambda \,,
\\[0.007ex]
\label{TB}
\nabla_\mu T^\alpha =& \; \partial_\mu T^\alpha + 
{\Gamma^\alpha}_{\mu\lambda} \, T^\lambda \,.
\end{align}
\end{subequations}
The third and fourth term on the right-hand side 
in Eq.~\eqref{condition} represent the spinor structure
contributions to the covariant derivative of the 
$\overline\gamma_\mu$ matrix.

The condition~\eqref{condition} defines 
the $\Gamma_\nu$ matrix up to a multiple of the 
unit matrix. 
In the vierbein formalism, we can represent
the $\overline\gamma^\nu$ matrices in terms of the 
vierbein $\wtilde\gamma^\mu$ matrices as follows,
\begin{subequations}
\label{ba}
\begin{align} 
\overline\gamma_\rho =& \; {b_\rho}^\alpha \, \wtilde\gamma_\alpha \,,
\qquad
\wtilde\gamma_\rho = {a^\alpha}_\rho \, \overline\gamma_\alpha \,,
\\[0.007ex]
\overline \gamma^\alpha =& \; {a^\alpha}_\rho \, \wtilde\gamma^\rho\,,
\qquad
\wtilde \gamma^\alpha = {b_\rho}^\alpha \, \overline\gamma^\rho \,.
\end{align} 
\end{subequations}
The metric is recovered as
\begin{subequations} 
\begin{align} 
\{ \overline \gamma_\rho, \overline \gamma_\sigma\} 
=& \; {b_\rho}^\alpha \, {b_\sigma}^\beta \, 
\{ \wtilde\gamma_\alpha , \wtilde\gamma_\beta \} 
= 2 \, \wtilde g_{\alpha \beta} \,
{b_\rho}^\alpha \, {b_\sigma}^\beta 
= 2 \, \overline g_{\rho \sigma} \,,
\\[0.007ex]
\{ \overline \gamma^\rho, \overline \gamma^\sigma\} 
=& \; 
{a^\rho}_\alpha \, {a^\sigma}_\beta \, 
\{ \wtilde\gamma^\alpha , \wtilde\gamma^\beta \} 
= 2 \, \wtilde g^{\alpha \beta} \,
{a^\rho}_\alpha \, {a^\sigma}_\beta \, 
= 2 \, \overline g^{\rho \sigma} \,.
\end{align}
\end{subequations} 
We note that the matrix with components $\overline g^{\rho \sigma}$
is the inverse of the matrix with components $\overline g_{\rho \sigma}$,
where the entries of $\wtilde g_{\alpha \beta}$ and 
$\wtilde g^{\alpha \beta}$ are identical.
It is possible to show, using rather lengthy 
algebra, that the following affine connection matrix,
\begin{equation} 
\label{solution}
\Gamma_\rho = - \frac{\ii}{4} \, \overline g_{\mu\alpha} \, 
\left( \frac{\partial {b_\nu}^\beta}{\partial x^\rho} \, 
{a^\alpha}_\beta - {\Gamma^\alpha}_{\nu \rho} \right) \, 
\overline\sigma^{\mu\nu} \,,
\end{equation}
with $\overline\sigma^{\mu\nu} = 
\frac{\ii}{2} \, [ \overline \gamma^\mu, \overline \gamma^\nu ]$,
fulfills Eq.~\eqref{condition} for a
general metric $\overline g^{\mu\nu}$.
Our result~\eqref{solution} differs from the result given 
in Eq.~(8) of Ref.~\cite{BrWh1957} 
in the correction of an obvious and in some sense
trivial typographical error. Namely, the expression
${\Gamma^\alpha}_{\nu l}$ in Eq.~(8) of Ref.~\cite{BrWh1957} 
should be replaced by the expression ${\Gamma^\alpha}_{\nu k}$,
(we have used $\rho$ for the corresponding subscript of 
$\Gamma_\rho$, not $\Gamma_k$ as in Ref.~\cite{BrWh1957}).
It is less trivial to see that a prefactor $1/4$ is missing from 
Eq.~(8) of Ref.~\cite{BrWh1957} and needs to be 
supplemented as given in Eq.~\eqref{solution}.
In order to clarify the matter, 
we should also point out that the additional imaginary unit in the prefactor 
is entirely due to our
different conventions for the $\gamma$ matrices and the 
flat-space metric which follow modern ``West--Coast''
conventions~\cite{ItZu1980,PeSc1995}.

It is rather lengthy but straightforward to show that
Eq.~\eqref{solution} solves Eq.~\eqref{condition}.
One needs to use Eqs.~\eqref{ba} and~\eqref{CGR} repeatedly,
and one needs to observe that the $b$ matrix is the inverse
of the $a$ matrix, i.e.,
${a^k}_\alpha \, {b_\rho}^\alpha = {\delta^k}_\rho$,
where $\delta$ is the Kronecker symbol.
Furthermore, the relation
${\Gamma^\beta}_{\sigma \rho} + {{\Gamma_\sigma}^\beta}_\rho =
{\overline g}^{\beta\alpha} \, \partial_\rho 
{\overline g}_{\alpha \sigma}$
is useful in intermediate steps of the calculation.
Here, ${{\Gamma_\sigma}^\beta}_\rho = 
g^{\beta\alpha} \, \Gamma_{\sigma \alpha \rho}$
with $\Gamma_{\sigma \alpha \rho}$ given in Eq.~\eqref{chr}.

Using the identity 
${\overline \sigma}^{\mu\nu} = \ii \, 
{\overline g}^{\mu\nu} - 
\ii {\overline \gamma}^\nu \, {\overline \gamma}^\mu$,
it is possible to rewrite Eq.~\eqref{solution} in a simpler form,
\begin{subequations} 
\begin{align} 
\label{SoMuGr}
\Gamma_\rho =& \; - \frac{\overline \gamma^\nu }{4} \, \left( 
\partial_\rho \overline \gamma_\nu 
- {\Gamma^\mu}_{\nu \rho} \, \overline \gamma_\mu \right) + 
{\mathcal A}_\rho \, \mathbbm{1}_{4 \times 4}\,,
\\[0.77ex]
{\mathcal A}_\rho =& \; \frac18 \,
\left[ 2 \left( \partial_\rho {b_\alpha}^\beta \right) \,
{a^\alpha}_\beta -
\left( \partial_\rho \overline g_{\alpha\beta} \right) \, 
\overline g^{\alpha\beta}  \right] \,.
\end{align}
\end{subequations} 
For a diagonal structure of the metric tensor (the only nonvanishing 
elements are the $\overline g_{\alpha\beta}$ with $\alpha=\beta$), 
with ${b_\alpha}^\beta = \sqrt{|\overline g_{\alpha\beta}|}$ and
${a^\alpha}_\beta = \sqrt{|\overline g^{\alpha\beta}|} =
1/\sqrt{|\overline g_{\alpha\beta}|}$,
the additional term ${\mathcal A}_\rho$ vanishes.
This is the case for the 
(generalized) Schwarzschild metric to be discussed below.
We have checked that, 
up to the term ${\mathcal A}_\rho$ and up to the 
conversion from ``East--Coast'' to ``West--Coast'' conventions, 
the result~\eqref{SoMuGr} is formally identical to the 
result previously given in Eq.~(9) of Ref.~\cite{SoMuGr1977}.
As a byproduct of our calculation of the ${\mathcal A}_\rho$,
we thus show that the two different results for the 
affine connection matrix given in 
Refs.~\cite{BrWh1957,SoMuGr1977} are equivalent
up to the term ${\mathcal A}_\rho \, \mathbbm{1}_{4 \times 4}$,
which is proportional to the unit matrix and 
not determined by the defining Eq.~\eqref{condition}.
For a diagonal metric, ${\mathcal A}_\rho$ vanishes.
To the best of our knowledge, the precise form of the 
term ${\mathcal A}_\rho$ has not yet been indicated in the 
literature.

Our construction of the spinor Lorentz transformation
in curved space [see Eq.~\eqref{Lloc}] 
follows ideas outlined in Ref.~\cite{Bo1975prd}.
However, our result for the covariant derivative of a spinor 
manifestly contains additional terms as compared to the 
result given in Eq.~(2.8) of Ref.~\cite{Bo1975prd}.
In particular, in view of the condition~\eqref{condition},
it is clear that the derivative terms
[the first terms in round brackets on the right-hand sides
of Eqs.~\eqref{solution} and~\eqref{SoMuGr}] 
are an essential contribution to the affine connection
matrix; these terms seem to be missing
in the vierbein formalism formulated in later steps of the 
derivation leading to Eq.~(2.8) of Ref.~\cite{Bo1975prd}.

%
%
\section{Schwarzschild--Type Metric}
\label{sec3}

\subsection{Radially Dependent Metrics}

In the following, we shall describe an important
application of the above formalism.
Namely, we shall discuss 
a generalization of the Schwarzschild metric, 
which describes, to good approximation,
the gravitational field of a planet, e.g.,
the Earth. In ``West--Coast'' conventions 
(where the sign of the timelike component is positive),
the Schwarzschild metric reads as 
follows~\cite{Sc1916,BrWh1957,Bo1975prd,SoMuGr1977},
\begin{align}
\label{schwarzschild}
{\overline g}_{\mu\nu} = & \;
{\rm diag}\left( \ee^\nu, -\ee^\lambda, -r^2, -r^2 \, \sin^2 \theta \right) 
\nonumber\\
=& \; {\rm diag}\left( u^2, -\frac{1}{u^2}, -r^2, -r^2 \, \sin^2 \theta \right) \,,
\\[0.007ex]
u^2 =& \; \ee^\nu = 
\ee^{-\lambda} = 1 - \frac{2 M_G}{r} \,,
\\[0.007ex]
r_s =& \; 2 M_G = 2 \, G \, M_P \,,
\end{align}
Here, the Schwarzschild radius is 
$r_s = 2 M_G = 2 G \, M_P / c^2 \approx 0.0089 \, {\rm m}$,
and $G$ is Newton's gravitational constant
(here, we supplement the factor $c^{-2}$ although we use units with 
$c=1$ in this article otherwise).
Furthermore, $M_P$ is the mass of the Earth (or, of the planet 
under consideration).  The invariant line element 
$\dd s^2$ in the Schwarzschild geometry is given by
\begin{align}
\label{ds2_ss}
\dd s^2 =& \; \left( 1 - \frac{r_s}{r} \right) \dd t^2 - 
\left( 1 - \frac{r_s}{r} \right)^{-1} \dd r^2 
\nonumber\\[0.007ex]
& \; - r^2 \, \left( \dd \theta^2 + \sin^\theta \, \dd \varphi^2 \right) \,.
\end{align}
The Schwarzschild metric is valid for a spherically symmetric geometry of
space. However, it has a problem.  Namely, as pointed out by 
Eddington~\cite{Ed1924}, for
the original Schwarzschild metric, the speed of light in the radial direction
is not equal to the speed of light in the transverse directions; the prefactor
in front of the ``angular'' term 
$r^2 \, \left( \dd \theta^2 + \sin^2 \theta \, \dd \varphi^2 \right)$
is not the same as the one in front of the 
``radial term'' proportional to $\dd r^2$.
This structure implies that one has to resort to a highly nonstandard 
representation of the Dirac algebra~\cite{SoMuGr1977} 
if one would like to separate the gravitationally coupled Dirac 
equation in the original form of the Schwarzschild metric~\eqref{schwarzschild}.

For example, without explicit mention, a representation of  the following 
form has apparently been used in Ref.~\cite{SoMuGr1977},
\begin{subequations}
\label{projective}
\begin{align}
\label{projective_a}
\wtilde\gamma^0 =& \;
\left( \begin{array}{cc} \mathbbm{1}_{2\times2} & 0 \\
0 & -\mathbbm{1}_{2\times2} \end{array} \right) \,,
\quad
\wtilde\gamma^1 =
\left( \begin{array}{cc} 0 & -\ii \, \mathbbm{1}_{2\times2} \\
-\ii \, \mathbbm{1}_{2\times2} & 0  \end{array} \right) \,,
\\[0.007ex]
\wtilde\gamma^2 = & \;
\left( \begin{array}{cc} 0 & -\sigma^3 \\ \sigma^3 & 0  \end{array} \right) \,,
\qquad
\wtilde\gamma^3 =
\left( \begin{array}{cc} 0 & \sigma^2 \\ -\sigma^2 & 0  \end{array} \right) \,,
\\[0.007ex]
\wtilde\gamma^5 =& \;
\ii \, \wtilde\gamma^0 \, \wtilde\gamma^1 \,
\wtilde\gamma^2 \, \wtilde\gamma^3
= \left( \begin{array}{cc} 0 & -\ii \sigma^1 \\ \ii \sigma^1 & 0
\end{array} \right) \,.
\end{align}
\end{subequations}
The authors of Ref.~\cite{SoMuGr1977} 
use the matrix $\wtilde\gamma^1$ for the ``radial''
part of the Dirac equation.
Specifically, near Eq.~(21) of Ref.~\cite{SoMuGr1977},
it is stated without further explanation that
a representation of the Dirac algebra is used where
${\widetilde \gamma}^1$ assumes a particularly simple form,
proportional to the expression given in Eq.~\eqref{projective_a}.
Indeed, such representations exist, as we show
in Eq.~\eqref{projective}, thus leading to a ramification
of the somewhat {\em ad hoc} statement made in Ref.~\cite{SoMuGr1977}.
It is easy to verify that the relations
$\{ \wtilde\gamma^\mu, \wtilde\gamma^\nu \} =
2 \, \wtilde g^{\mu\nu} = 2\, {\rm diag}(1,-1,-1,-1)$
are fulfilled. Here, the $\vec\sigma = (\sigma^1, \sigma^2, \sigma^3)$ are the
($2 \otimes 2$)--Pauli spin matrices,
and $\mathbbm{1}_{2\times2}$ denotes the
($2 \otimes 2$) unit matrix.

%
%
\subsection{Eddington's Reparameterization}

In Sec.~43 of Chap.~3 of Ref.~\cite{Ed1924}, Eddington has pointed out that a
coordinate transformation exists which converts the 
Schwarzschild metric into spatially isotropic form.
It reads as follows,
\begin{equation}
r = r_1 \left( 1 + \frac{r_s}{4 r_1} \right)^2 \,,
\quad
r_1 = \frac{r}{2} - \frac{r_s}{4} + 
\sqrt{\frac{r}{4} \, \left( r-r_s\right)} \,.
\end{equation}
After this transformation, the invariant 
line element~\eqref{ds2_ss} becomes
\begin{align}
\dd s^2 =& \; 
\frac{\left(4 r_1 - r_s\right)^2}%
{\left(4 r_1 + r_s\right)^2} \, \dd t^2 
\\[0.77ex]
& \; -\left( 1 + \frac{r_s}{4 r_1} \right)^4 \;
\left( \dd r_1^2 + r_1^2 \dd \theta^2 + 
r_1^2 \, \sin^2 \theta \, \dd \varphi^2 \right) \,.
\nonumber
\end{align}
Using this isotropic form of the metric, we can now 
transform the spatial part to Cartesian coordinates,
\begin{align}
\dd s^2 =& \; \frac{\left(4 r_1 - r_s\right)^2}%
{\left(4 r_1 + r_s\right)^2} \dd t^2
-\left( 1 + \frac{r_s}{4 r_1} \right)^4 
\left( \dd x_1^2 + \dd y_1^2 + \dd z_1^2 \right),
\end{align}
where $x_1 = r_1 \, \sin\theta \, \cos\varphi$,
$y_1 = r_1 \, \sin\theta \, \sin\varphi$, 
and $z_1 = r_1 \, \cos\theta$.
We now redefine 
\begin{subequations}
\begin{align}
r_1 \to r, &\; \quad x_1 \to x, \quad y_1 \to y, \quad z_1 \to z,
\\[0.117ex]
r =& \; \sqrt{x^2 + y^2 + z^2} \,.
\end{align}
\end{subequations}
Furthermore, we define $w(r)$ and $v(r)$
as follows,
\begin{equation}
\label{vwspec}
w(r) = \frac{4 r - r_s}{4 r + r_s} \,,
\quad
v(r) = \left( 1 + \frac{r_s}{4 r} \right)^2 \,.
\end{equation}
The transformed (according to Ref.~\cite{Ed1924}) 
Schwarzschild metric can now be written
in the following form,
\begin{equation}
\label{vw}
{\overline g}_{\mu\nu} = 
{\rm diag}\left( w^2(r), -v^2(r), -v^2(r), -v^2(r) \right) \,.
\end{equation}
The considerations below are valid for a general
form~\eqref{vw} of the metric and not tied to the 
specific form given in Eq.~\eqref{vwspec}.
The vierbein coefficients are given as
\begin{align}
{b_0}^\beta =& \; {b_\beta}^0 = {\delta_0}^\beta \, w(r) \,,
\qquad
{b_i}^j = {\delta_i}^j \, v(r) \,,
\\[0.77ex]
{a^\alpha}_0 =& \; {a^0}_\alpha = 
\frac{{\delta^0}_\alpha}{w(r)} \,,
\qquad
{a_i}^j = \frac{{\delta_i}^j}{v(r)} \,,
\end{align}
where $i,j=1,2,3$ and ${\delta_\alpha}^\beta$ and 
${\delta^\alpha}_\beta$ are Kronecker symbols
(i.e., equal to one if the indices are equal, 
otherwise zero).

With these coefficients, using computer algebra~\cite{Wo1988},
it is easy to evaluate all Christoffel symbols and to 
establish that 
\color{black}
\begin{subequations}
\begin{align}
{\overline \gamma}^0 \, 
{\overline \gamma}^\mu \, 
\Gamma_\mu =& \;
- \frac{\wtilde \gamma^0 \, 
\vec{\wtilde \gamma} \cdot \hat r}{v(r) \, w(r)} \, G(r) \,,
\\[2ex]
G(r) =& \; \frac{ 2 \, v'(r) \, w(r) + v(r) \, w'(r)}{2 \, v(r) \,w(r)} \,.
\end{align}
\end{subequations}
\color{black}
This result has been verified by us both using the 
representation~\eqref{solution} as well as the representation
given in~\eqref{SoMuGr}, for the metric~\eqref{vw}.

%
%
\subsection{Reduction to Radial Equation}

In our further analysis,
we use the Hamiltonian form of the gravitationally 
coupled Dirac equation,
\begin{equation}
\label{noncov}
\ii \left( {\overline \gamma}^0 \right)^2\,\partial_t \psi = 
\left( \vec{\overline \alpha} \cdot \vec p + 
\ii \, {\overline \gamma}^0 \, {\overline \gamma}^\mu \, \Gamma_\mu + 
{\overline \gamma}^0 \, m \right) \, \psi \,,
\end{equation}
where $\vec p = -\ii \, \partial/\partial \vec r$ 
with $\vec r = (x,y,z)$, 
$\overline \alpha^i = \overline \gamma^0 \, \overline \gamma^i$,
and the expression ``Hamiltonian form'' is 
used in analogy with flat-space.
Namely, in flat space, the expression on the left-hand side 
simply represents the ``noncovariant time-evolution 
operator'' because $\left({\overline \gamma}^0 \right)^2 \to
\left({\tilde \gamma}^0 \right)^2 = \mathbbm{1}_{4 \times 4}$
and $H \to \ii \, \partial_t$ 
(see Refs.~\cite{Ro1961,SwDr1991a,SwDr1991b,SwDr1991c}).
In curved space, with the metric given in Eq.~\eqref{vw},
we have $\left({\overline \gamma}^0 \right)^2 = 
\tfrac{1}{w^2(r)} \, \mathbbm{1}_{4 \times 4}$.

In the following, we use the Dirac matrices in the Dirac 
representation,
\begin{subequations}
\label{dirac_rep}
\begin{align}
\wtilde\gamma^0 =& \;
\left( \begin{array}{cc} \mathbbm{1}_{2\times2} & 0 \\
0 & -\mathbbm{1}_{2\times2} \end{array} \right) \,,
\quad
\wtilde\gamma^1 =
\left( \begin{array}{cc} 0 & \sigma^1 \\ -\sigma^1 & 0  \end{array} \right) \,,
\\[0.007ex]
\wtilde\gamma^2 = & \;
\left( \begin{array}{cc} 0 & \sigma^2 \\ -\sigma^2 & 0  \end{array} \right) \,,
\qquad
\wtilde\gamma^3 =
\left( \begin{array}{cc} 0 & \sigma^3 \\ -\sigma^3 & 0  \end{array} \right) \,,
\\[0.007ex]
\wtilde\gamma^5 =& \;
\ii \, \wtilde\gamma^0 \, \wtilde\gamma^1 \,
\wtilde\gamma^2 \, \wtilde\gamma^3
= \left( \begin{array}{cc} 0 & \mathbbm{1}_{2\times2} \\ 
\mathbbm{1}_{2\times2} & 0 \end{array} \right) \,.
\end{align}
\end{subequations}
In the form~\eqref{noncov}, the gravitationally coupled Dirac equation
allows a solution of the standard form~\cite{Ro1961,SwDr1991a,SwDr1991b,SwDr1991c}
\begin{equation}
\label{psi_stat}
\psi = \left( \begin{array}{c}
f(r) \; \chi_{\varkappa\mu}(\theta, \varphi) \\[0.77ex]
\ii g(r) \; \chi_{-\varkappa\mu}(\theta, \varphi)
\end{array} \right) \, \exp(-\ii \, E \, t) \,,
\end{equation}
where the $ \chi_{\varkappa\mu}(\theta, \varphi)$ 
[sometimes denoted as $ \chi_\varkappa^\mu(\theta, \varphi)$] are the 
standard spin-angular functions~\cite{Ro1961,SwDr1991a,SwDr1991b,SwDr1991c}.
They have the property
\begin{equation}
(\vec \sigma \cdot \vec L + 1) \, \chi_{\varkappa\mu}(\theta, \varphi) = 
-\varkappa \, \chi_{\varkappa\mu}(\theta, \varphi) \,.
\end{equation}
We recall that the eigenvalues 
of the operator $K = \vec \sigma \cdot \vec L + 1$ 
are $-\varkappa$ [see the text following
Eq.~(2.9) of Ref.~\cite{SwDr1991a}].
In the text following Eq.~(18) of 
Ref.~\cite{SoMuGr1977}, the eigenvalues is assumed
to be $+\varkappa$, an apparent typographical error.
It is extremely instructive to write the 
Hamiltonian form~\eqref{noncov}, 
using the ansatz~\eqref{psi_stat}, in terms of 
$(2 \otimes 2)$ spin matrices,
\color{black}
\begin{widetext}
\begin{align}
\frac{\ii \partial_t\psi(\vec r)}{w^2(r)} =& \;
\left( \begin{array}{cc}
\dfrac{m}{w(r)} & \dfrac{\vec \sigma \cdot \hat r}{v(r) \, w(r)} \,
\left( - \ii \, \dfrac{\partial}{\partial r} +
\ii \, \dfrac{\vec \sigma \cdot \vec L}{r} - \ii \, G(r) \right) \\[0.33ex]
\dfrac{\vec \sigma \cdot \hat r}{v(r) \, w(r)} \,
\left( - \ii \, \dfrac{\partial}{\partial r} +
\ii \, \dfrac{\vec \sigma \cdot \vec L}{r} - \ii \, G(r) \right) &
-\dfrac{m}{w(r)} \\
\end{array} \right) \,
\left( \begin{array}{c}
f(r) \, \chi_{\varkappa \, \mu}(\hat r) \\[2ex]
\ii g(r) \, \chi_{-\varkappa \, \mu}(\hat r) \\
\end{array} \right)
\nonumber\\[2ex]
=& \;
\left( \begin{array}{c}
\left[ \dfrac{1}{v(r) \, w(r)} \left( -\dfrac{\partial}{\partial r} +
\dfrac{1}{r} \, \left( \varkappa - 1\right) - G(r)
\right) \, g(r) +  \dfrac{m}{w(r)} \, f(r) \right] \,
\chi_{\varkappa \, \mu}(\hat r) \\[3ex]
\ii \, \left[ \dfrac{1}{v(r) \, w(r)} \left( \dfrac{\partial}{\partial r} +
\dfrac{1}{r} \, \left( \varkappa + 1\right) + G(r)
\right) \, f(r) -  \dfrac{m}{w(r)} \, g(r) \right] \,
\chi_{-\varkappa \, \mu}(\hat r)
\end{array} \right) =
\frac{E}{w^2(r)} \, \left( \begin{array}{c}
f(r) \, \chi_{\varkappa \, \mu}(\hat r) \\[3ex]
\ii \, g(r) \, \chi_{-\varkappa \, \mu}(\hat r)
\end{array} \right)\,.
\end{align}
\end{widetext}
\color{black}
Here, we have used the relation 
$\vec\sigma \cdot \hat r \, \chi_{\varkappa\mu}(\hat r) = 
- \chi_{-\varkappa\mu}(\hat r)$, 
which can be found in Eq.~(7.2.5.23) of Ref.~\cite{VaMoKh1988},
where $\hat r = \vec r /|\vec r|$ is the position 
unit vector. The radial equations are thus given as
\color{black}
\begin{subequations}
\label{rad}
\begin{align}
\left( \dfrac{\partial}{\partial r} +
\dfrac{1 - \varkappa}{r} + G(r) \right) \, g(r)
=& \; v(r) \, \left(m - \frac{E}{w(r)}\right) \, f(r) \,,
\\[2ex]
\left( \dfrac{\partial}{\partial r} +
\dfrac{1 + \varkappa}{r} +
G(r) \right) \, f(r) =& \; v(r) \,
\left(m + \frac{E}{w(r)}\right) \, g(r) \,.
\end{align}
\end{subequations}
\color{black}
An important symmetry property of Eq.~\eqref{rad} is
given by its invariance under the simultaneous
replacements
\begin{equation}
\label{sym}
E \leftrightarrow -E,
\qquad f(r) \leftrightarrow g(r) \,,
\qquad \varkappa \leftrightarrow -\varkappa \,.
\end{equation}
So, if $E$ is an eigenvalue of the gravitationally coupled
Dirac equation, so is $-E$. Invoking the reinterpretation
principle~\cite{St1941,St1942,Fe1949}
and interpreting the negative energy $-E < 0$
as $+E > 0$ for antiparticles (which propagate
``into the past''), we find that the spectrum of the gravitationally coupled
Dirac Hamiltonian is exactly the same for particles and antiparticles.
This important finding is true for any space-time metric
of the form given in Eq.~\eqref{vw} and 
not necessarily tied to the Schwarzschild geometry.

Let us now establish the connection to the 
flat-space result given in Ref.~\cite{SwDr1991a}.
Specifically, Eqs.~(2.12a) and Eq.~(2.12b) of Ref.~\cite{SwDr1991a},
using the identity $r^{-1} \partial_r (r \, h(r)) =
\partial_r h(r) + r^{-1} \, h(r)$,
for $\hbar = c = 1$ and $Z \alpha \to 0$, can be written as
\begin{subequations}
\begin{align}
\left( \frac{\partial}{\partial r} + \frac{1-\varkappa}{r} \right) g(r) =& \; 
(m - E) \, f(r) \,,
\\[0.007ex]
\left( \frac{\partial}{\partial r} + \frac{1+\varkappa}{r} \right) f(r)) =& \;
(m + E) \, g(r) \,,
\end{align}
\end{subequations}
where we have used the form~\eqref{psi_stat} for the 
wave function. These equations therefore become identical to our 
Eq.~\eqref{rad} in the limit $v(r) \to 1$, $w(r) \to 1$.

The symmetry property~\eqref{sym} is physically 
tied to the reinterpretation principle which is very well
known in the particle physics 
community~\cite{BjDr1964,BjDr1965,ItZu1980,PeSc1995} but less well
known in the general relativity community.
Some remarks are therefore in order.
We consider a space-time interval $\Delta x = (\Delta t, \Delta \vec r)$,
and a scalar product $k \cdot \Delta x =
|E| \, \Delta t - \vec k \cdot \Delta\vec r $ (with $\Delta t > 0$).
The antiparticle amplitude 
$\exp(\ii \, k \cdot \Delta x)$ then is proportional to 
\begin{equation}
\ee^{\ii |E| \, \Delta t - \ii \vec k \cdot \Delta\vec r } \to
\ee^{-\ii |E| \, (-\Delta t) + \ii \vec k \cdot (-\Delta\vec r) } 
\end{equation}
where $-\Delta t > 0$ and one can thus reinterpret the 
antiparticle trajectory, initially propagating 
``into the past'' (advanced contribution to the 
Feynman propagator) and along the distance interval $\Delta \vec r$
(``from point $a$ to point $b$''), 
as a positive-energy trajectory with energy $+E$,
covering the inverse space-time interval $(-\Delta t > 0, -\Delta \vec r)$
i.e., propagating into the future with 
four-momentum $(|E|, \vec k)$, i.e., ``from point $b$ to point $a$.''
Applied to gravitational interactions, the currently
available accepted interpretation based on particle
physics principles~\cite{BjDr1964,BjDr1965,ItZu1980,PeSc1995}
therefore dictates that ``an antiparticle falls upward in the gravitational
field, but backward in time, and with the same modulus of the 
kinetic energy as the corresponding particle.''
Therefore, after reinterpretation,
the formalism of the gravitationally 
coupled Dirac equation predicts that
antiparticles and particles receive exactly the same
energy perturbations in a gravitational field,
at least within space-time geometries that have the 
general form~\eqref{vw}.
This important 
result generalizes Eq.~\eqref{eqEP} to the relativistic
quantum domain.

%
%
\section{Conclusions}

In this article, we reexamine the 
gravitationally coupled Dirac equation in Sec.~\ref{sec2},
explaining a number of aspects of the derivation in
greater detail.
In particular, we show that the condition~\eqref{condition}
follows naturally as a consequence of the 
fundamental anticommutator property of the 
curved-space Dirac matrices~\eqref{gmunu},
together with the known fact that the 
covariant derivative of the metric tensor has to 
vanish~\cite{MiThWh1973}. Under these assumptions,
the covariant derivative of the curved-space
Dirac matrix $\overline\gamma^\mu(x)$ also has to vanish,
and the condition~\eqref{condition} 
follows as a consequence of the ansatz~\eqref{ansatzG}
for the covariant derivative of a spinor,
together with the fundamental commutator property~\eqref{commutes}.
The symmetrization of the 
covariant action of the Dirac field given in Eq.~\eqref{symS} 
then becomes possible, in analogy to the flat-space
action Eq.~\eqref{S0}.
Furthermore, under a proper definition of the 
local spinor Lorentz transformation~\eqref{Lloc}, 
expressed in space-time coordinates, the 
local Dirac matrices $\overline\gamma^\mu = \overline\gamma^\mu(x)$ are
shape-invariant, as shown in Eq.~\eqref{ensure}.
For a general metric ${\overline g}^{\mu\nu} = {\overline g}^{\mu\nu}(x)$,
we find the vierbein representation~\eqref{solution} of the 
affine connection matrices $\Gamma_\rho = \Gamma_\rho(x)$
which differs from the result given previously 
in Eq.~(8) of Ref.~\cite{BrWh1957} by a factor~$1/4$. 
With the additional prefactor,
the result given in Eq.~\eqref{solution} 
then is in agreement with the result for the 
affine connection matrices given in 
Eq.~(9) of Ref.~\cite{SoMuGr1977}.
In ``West--Coast'' conventions for the metric,
the gravitationally Dirac equation reads as
$\left( \ii \, \overline\gamma^\mu \, \nabla_\mu - m \right) \psi(x) = 0$
[see Eq.~\eqref{gravdirac}],
as opposed to the ``East--Coast'' 
form $\left( \overline\gamma^\mu \, \nabla_\mu + m \right) \psi(x) = 0$
[see Refs.~\cite{BrWh1957,SoMuGr1977}].

The gravitationally coupled radial Dirac equation
given in~\eqref{rad} for a Schwarzschild-type metric~\eqref{vw}
describes the coupling of a particle 
(and corresponding antiparticle) to the gravitational
field of a planet. Our Eq.~\eqref{rad}, in appropriate limits,
is in agreement with the fundamental properties 
of upper and lower components describing particles
and antiparticles at rest ($E \to m$ and $E \to -m$),
if we additionally consider the limit 
of flat space-time [$v(r) \to 1$, $w(r) \to 1$]. 
This limit is explored easily, starting from 
Eq.~(2.12) of Ref.~\cite{SwDr1991a}.

The symmetry $E \leftrightarrow -E$, $f \leftrightarrow g$,
$\varkappa \leftrightarrow -\varkappa$
given in Eq.~\eqref{sym} implies that the quantum states
of spin-$1/2$ antiparticles, in the 
gravitational field of the Earth, have exactly the same 
spectrum as those of the corresponding particles, 
including all relativistic corrections of motion.
Therefore, this statement also holds for superpositions of quantum states.
including those which describe a wave packet 
evolving along a classical trajectory (these states would
other be known as ``coherent'' or ``Glauber'' states).
Our quantum-theory based findings go beyond the simple statement that 
particle and antiparticle trajectories in 
curved space-time are the same on the 
classical level [see Eq.~\eqref{eqEP}].

Finally, let us include a remark regarding the validity of the 
gravitationally coupled Dirac equation for antiparticles.
One might contemplate if antiparticles should be
described by a different equation in the context of gravity than particles, but
by the same equation in the context of electromagnetism (electromagnetically
coupled Dirac equation).
In this case, the gravitationally coupled Dirac equation~\eqref{gravdirac}
would only describe particles, not antiparticles,
even if it admits negative-energy solutions.
However, in this case one gets into trouble 
in the limit of a vanishing gravitational interaction,
in which case space becomes flat. This is because the
Dirac equation is known to describe antiparticles 
very well in this limit~\cite{BjDr1964,BjDr1965,ItZu1980,PeSc1995}.
At least, this concept is used in all perturbative QED 
calculations, including the notoriously difficult 
bound-state problems~\cite{MoPlSo1998}.
If we conjecture that the flat-space limit is smooth, then the gravitationally coupled
Dirac equation~\eqref{gravdirac} must remain valid for both particles and 
antiparticles. Brill and Wheeler~\cite{BrWh1957}, 
Boulware~\cite{Bo1975prd}, and
Soffel, M\"{u}ller, and Greiner~\cite{SoMuGr1977}
all used the same methods for deriving the coupled Dirac equation 
(we here attempt to resolve some discrepancies found in the literature
regarding the final steps in the derivation).
The gravitationally coupled 
Dirac equation~\eqref{gravdirac} involves $(4\otimes 4)$ matrices
and allows for two positive-energy solutions,
which are naturally interpreted as particles,
and two negative-energy solutions, which are naturally
interpreted as antiparticles, according to usual practice in particle 
physics~\cite{BjDr1964,BjDr1965,ItZu1980,PeSc1995}.
We may thus assume that  the gravitationally coupled
Dirac equation given in Eq.~\eqref{gravdirac}
should be valid for particles and antiparticles,
simultaneously. We know for sure that the corresponding variant of the equation
in flat space describes particles and antiparticles
simultaneously, as described in detail, e.g., in Chap.~2 of Ref.~\cite{ItZu1980}.

Despite our theoretical considerations,
it is still of utmost value to the 
scientific community to carry out the planned 
antimatter gravity experiments~\cite{ALPHA,ATHENA,ASACUSA,ATRAP,AGELOI,Ke2008}.
We conclude with two remarks:
(i)~The ``inertial mass term'' in the sense
of the equivalence principle enters the
Schr\"{o}dinger equation for 
free and bound electrons, e.g., for a bound electron in 
a hydrogen atom. According to experimental 
evidence, inertial and gravitational mass are the 
the same for atoms (such as atomic hydrogen), which is composed of 
spin-$1/2$ particles (electrons and protons),
therefore, the action principle~$\delta S = 0$
[see Eq.~\eqref{Sdirac}] provides for a solid basis of the 
discussion of relativistic quantum effects in gravitational 
coupling, with the $m$ term entering the 
equation being equal to the inertial (gravitational) mass. 
(ii) Our investigations 
suggest that any conceivable differences of 
the gravitational coupling of particles and antiparticles
should be assigned to a ``fifth force,''
not to any conceivable ``modifications of 
the gravitational mass'' of antiparticles versus particles.

%
%
\section*{Acknowledgments}

The author acknowledges helpful discussions with 
Zolt\'{a}n Tr\'{o}cs\'{a}nyi and Istv\'{a}n N\'{a}ndori, and support by 
the National Science Foundation (Grant No.~PHY--1068547) and
by the National Institute of Standards and Technology 
(precision measurement grant).


\begin{thebibliography}{10}

\bibitem{GaEtAl2008}
\relax{G. Gabrielse {\em et al.} [ATRAP Collaboration]}, Phys. Rev. Lett. {\bf
  100},  113001  (2008).

\bibitem{AmEtAl2011}
\relax{C. Amole {\em et al.} [ALPHA Collaboration]}, Nature (London) {\bf 483},
   439  (2012).

\bibitem{ALPHA}
ALPHA Collaboration {\em (Antihydrogen Laser Physics Apparatus)}, see the URL
  http://alpha-new.web.cern.ch.

\bibitem{ATHENA}
ATHENA Collaboration {\em (ATHENA Antihydrogen Apparatus)}, see the URL
  http://athena.web.cern.ch/.

\bibitem{ASACUSA}
ASACUSA Collaboration {\em (Atomic Spectroscopy And Collisions Using Slow
  Antiprotons)}, see the URL http://asacusa.web.cern.ch/.

\bibitem{ATRAP}
ATRAP Collaboration {\em (Antihydrogen trap)}, see the URL
  http://gabrielse.physics.harvard.edu/.

\bibitem{AGELOI}
\relax{A. D. Cronin {\em et al.} [AGE Collaboration]}, Letter of Intent:
  Antimatter Gravity Experiment (AGE) at Fermilab (2009), available at the URL
  http://www.fnal.gov/\allowbreak{}directorate/\allowbreak{}program\_planning/%
  \allowbreak{}Mar2009PACPublic/\allowbreak{}AGELOIFeb2009.pdf; see also the
  URL
  http://www.phy.duke.edu/\~{}phillips/\allowbreak{}gravity/\allowbreak{}frameIndex.html.

\bibitem{Ke2008}
\relax{A. Kellerbauer {\em et al.} [AEGIS Proto-Collaboration]}, Nucl. Instrum.
  Methods Phys. Res. B {\bf 266},  351  (2008).

\bibitem{ALPHACOILS}
see the URL
  http://www.quantumdiaries.org/\allowbreak{}2011/06/06/\allowbreak{}%
  trapping-antimatter-with-magnets/.

\bibitem{MiThWh1973}
C.~W. Misner, K.~S. Thorne, and J. Wheeler, {\em \relax{Gravitation}} (W. H.
  Freeman, New York, 1973).

\bibitem{Ko1996}
M. Kowitt, Int. J. Theor. Phys. {\bf 35},  605  (1996).

\bibitem{BrWh1957}
D.~R. Brill and J.~A. Wheeler, Rev. Mod. Phys. {\bf 29},  465  (1957).

\bibitem{Bo1975prd}
D.~G. Boulware, Phys. Rev. D {\bf 12},  350  (1975).

\bibitem{SoMuGr1977}
M. Soffel, B. M\"{u}ller, and W. Greiner, J. Phys. A {\bf 10},  551  (1977).

\bibitem{Di1928a}
P.~A.~M. Dirac, Proc. Roy. Soc. London, Ser. A {\bf 117},  610  (1928).

\bibitem{Di1928b}
P.~A.~M. Dirac, Proc. Roy. Soc. London, Ser. A {\bf 118},  351  (1928).

\bibitem{BjDr1964}
J.~D. Bjorken and S.~D. Drell, {\em \relax{Relativistic Quantum Mechanics}}
  (McGraw-Hill, New York, 1964).

\bibitem{BjDr1965}
J.~D. Bjorken and S.~D. Drell, {\em \relax{Relativistic Quantum Fields}}
  (McGraw-Hill, New York, 1965).

\bibitem{ItZu1980}
C. Itzykson and J.~B. Zuber, {\em \relax{Quantum Field Theory}} (McGraw-Hill,
  New York, 1980).

\bibitem{PeSc1995}
M.~E. Peskin and D.~V. Schroeder, {\em \relax{An Introduction to Quantum Field
  Theory}} (Perseus, Cambridge, Massachusetts, 1995).

\bibitem{ShCa1991}
G.~V. Shishkin and W.~D. Cabos, J. Math. Phys. {\bf 33},  914  (1991).

\bibitem{ShVa1992}
G.~V. Shishkin and V.~M. Villalba, J. Math. Phys. {\bf 33},  4037  (1992).

\bibitem{Sc1916}
K. Schwarzschild, Sitzungsberichte d. K. Preuss. Akad. d. Wiss. {\bf 7},  189
  (1916).

\bibitem{Ed1924}
A.~S. Eddington, {\em \relax{The Mathematical Theory of Relativity}} (Cambridge
  University Press, Cambridge, England, 1924).

\bibitem{Wo1988}
S. Wolfram, {\em \relax{Mathematica-A System for Doing Mathematics by
  Computer}} (Addison-Wesley, Reading, MA, 1988).

\bibitem{Ro1961}
M.~E. Rose, {\em \relax{Relativistic Electron Theory}} (J. Wiley \& Sons, New
  York, NY, 1961).

\bibitem{SwDr1991a}
R.~A. Swainson and G.~W.~F. Drake, J. Phys. A {\bf 24},  79  (1991).

\bibitem{SwDr1991b}
R.~A. Swainson and G.~W.~F. Drake, J. Phys. A {\bf 24},  95  (1991).

\bibitem{SwDr1991c}
R.~A. Swainson and G.~W.~F. Drake, J. Phys. A {\bf 24},  1801  (1991).

\bibitem{VaMoKh1988}
D.~A. Varshalovich, A.~N. Moskalev, and V.~K. Khersonskii, {\em \relax{Quantum
  Theory of Angular Momentum}} (World Scientific, Singapore, 1988).

\bibitem{St1941}
E. St\"{u}ckelberg, Helv. Phys. Acta {\bf 14},  588  (1941).

\bibitem{St1942}
E. St\"{u}ckelberg, Helv. Phys. Acta {\bf 15},  23  (1942).

\bibitem{Fe1949}
R.~P. Feynman, Phys. Rev. {\bf 76},  769  (1949).

\bibitem{MoPlSo1998}
P.~J. Mohr, G. Plunien, and G. Soff, Phys. Rep. {\bf 293},  227  (1998).

\end{thebibliography}
\end{document}